\documentclass[aps,prl,reprint,superscriptaddress,showpacs,showkeys]
{revtex4-1}
\usepackage{graphicx}
\usepackage{array}
\usepackage[caption=false]{subfig}
\usepackage{amssymb}
\usepackage[utf8]{inputenc}

\begin{document}


\title{Multiple Galactic Sources with Emission Above 56 TeV Detected by HAWC}



\author{A.U.~Abeysekara}
\affiliation{Department of Physics and Astronomy, University of Utah, Salt Lake City, UT, USA }
\author{A.~Albert}
\affiliation{Physics Division, Los Alamos National Laboratory, Los Alamos, NM, USA }
\author{R.~Alfaro}
\affiliation{Instituto de F\'{i}sica, Universidad Nacional Autónoma de México, Ciudad de Mexico, Mexico }
\author{J.R.~Angeles Camacho}
\affiliation{Instituto de F\'{i}sica, Universidad Nacional Autónoma de México, Ciudad de Mexico, Mexico }
\author{J.C.~Arteaga-Velázquez}
\affiliation{Universidad Michoacana de San Nicolás de Hidalgo, Morelia, Mexico }
\author{K.P.~Arunbabu}
\affiliation{Instituto de Geof\'{i}sica, Universidad Nacional Autónoma de México, Ciudad de Mexico, Mexico }
\author{D.~Avila Rojas}
\affiliation{Instituto de F\'{i}sica, Universidad Nacional Autónoma de México, Ciudad de Mexico, Mexico }
\author{H.A.~Ayala Solares}
\affiliation{Department of Physics, Pennsylvania State University, University Park, PA, USA }
\author{V.~Baghmanyan}
\affiliation{Institute of Nuclear Physics Polish Academy of Sciences, PL-31342 IFJ-PAN, Krakow, Poland }
\author{E.~Belmont-Moreno}
\affiliation{Instituto de F\'{i}sica, Universidad Nacional Autónoma de México, Ciudad de Mexico, Mexico }
% ...

\collaboration{HAWC Collaboration}

\date{\today}

\begin{abstract}
We present the first catalog of gamma-ray sources emitting above 56 and 100 TeV with data from the High Altitude Water Cherenkov (HAWC) Observatory, a wide field-of-view observatory capable of detecting gamma rays up to a few hundred TeV. Nine sources are observed above 56 TeV, all of which are likely Galactic in origin. Three sources continue emitting past 100 TeV, making this the highest-energy gamma-ray source catalog to date. We report the integral flux of each of these objects. We also report spectra for three highest-energy sources and discuss the possibility that they are PeVatrons.
\end{abstract}


\maketitle

\section{Introduction}
The all-particle cosmic-ray (CR) spectrum contains a break called the ``knee" at $\sim$1 PeV~\cite{pdg}. CRs are expected to be Galactic in origin up to at least this point. Identifying sources that accelerate particles to this energy (``PeVatrons") can help us understand this feature.

The question of which source classes can be PeVatrons is still open. Supernova remnants (SNRs) have traditionally been suggested as the most plausible candidates~\cite{Aharonian2013}. However, theories of CR acceleration in SNRs begin to encounter problems at a few hundred TeV~\cite{Gabici2016,Bell2013}. Alternative PeVatron candidates include young massive star clusters~\cite{Aharonian} and supermassive black holes~\cite{Abramowski2016}. The only previously reported PeVatron (the Galactic center region, by the H.E.S.S. Collaboration~\cite{Abramowski2016}) has been hypothesized to be the latter. This source does not have a high enough current rate of particle acceleration to provide a sizable contribution to Galactic CRs but could have been more active in the past.

Since CRs are charged, they bend in magnetic fields on their way to Earth and are difficult to trace back to their sources. Neutral gamma rays can instead be used to probe PeVatrons. When CRs interact with their environment (the interstellar medium, an ambient photon, or the gas/plasma of an SNR), the particles created include neutral pions. Each $\pi^0$ decays to two gamma rays. For a PeV CR, the gamma ray is approximately one order of magnitude less energetic~\cite{Hinton2009}. A source with a hard gamma-ray spectrum (power-law index 2-2.4) extending to 100 TeV without an apparent spectral cutoff would be a clear signature of a PeVatron~\cite{Aharonian2013}.

Charged pions, which are also created in these hadronic interactions, produce neutrinos. A sub-dominant ($<$14$\%$) fraction of the IceCube astrophysical neutrinos~\cite{Aartsen2013,Aartsen2014} could be Galactic in origin and also associated with PeVatrons~\cite{Aartsen2017}. Gamma-ray and neutrino measurements could be used together to probe PeVatrons.

Gamma rays are also produced via leptonic processes; at TeV energies inverse Compton (IC) scattering is the dominant mechanism. Above a few tens of TeV, the leptonic component of gamma-ray emission becomes suppressed due to Klein-Nishina effects. This results in an energy-dependent spectral index~\cite{Moderski2005}. Observations above 50 TeV are essential in identifying PeVatron candidates. If the spectrum of a source exhibits significant curvature, it is more likely to be dominated by leptonic emission.

Using data from the High Altitude Water Cherenkov (HAWC) Observatory~\cite{Smith2015,Abeysekara2018}, we present the highest-energy gamma-ray sky survey ever performed. HAWC is a wide field-of-view experiment that has unprecedented sensitivity at the highest photon energies~\cite{Abeysekara2017a} and excellent sensitivity to extended sources (the integral flux $>$ 2 TeV is $\sim$10$^{-13}$ cm$^{-2}$ sec$^{-1}$ for a source extent of 0.5$^{\circ}$~\cite{Abeysekara2013}) . These characteristics are crucial for detecting PeVatrons.

HAWC observations can also be used to look for signatures of Lorentz Invariance violation (LIV). In some extensions of the Standard Model, the highest-energy photons decay quickly, with the decay probability near 100$\%$ over astrophysical distance scales~\cite{Martinez-Huerta2017a,MartinezHuerta2019}. Therefore, the existence of photons from astrophysical sources above 100 TeV constrains the linear effect of LIV to be $>$ 9.6$\times10^{29}$ eV (78 times the Planck mass)~\cite{Martinez-Huerta2017}.  This paper focuses on the evidence of the sources detected by HAWC with $>$ 100 TeV photons. Further analysis of the highest-energy photons and their LIV implications will be discussed in a future publication.  

\section{Analysis method}
HAWC uses two recently developed energy estimation algorithms which have been used to identify $>$ 56 TeV gamma rays from the Crab Nebula~\cite{crab2018}. In this work, we use the ``ground parameter'' method. Throughout this paper, $\hat{E}$ refers to estimated energy.

The analysis is performed in three steps: source identification, localization, and spectral fits. The data were collected between June 2015 and July 2018 (total livetime: 1038.8 days). The background rejection, event binning, and likelihood framework~\cite{Vianello2015} as described in \cite{crab2018} are used to create $\sqrt{TS}$ (test statistic, defined as \mbox{-2ln($\mathcal{L}_{1}/\mathcal{L}_{0}$}), see Supplemental Material~\cite{supp}) maps of the high-energy sky above two $\hat{E}$ thresholds: 56 TeV and 100 TeV. Sources in these maps are identified by applying the same technique used for the 2HWC catalog~\cite{Abeysekara2017b}. The declination range searched is -20$^{\circ}$ to 60$^{\circ}$. The maps are made assuming a power-law spectrum with an index of -2.0 and three different source morphology assumptions (point source as well as disks of radii 0.5$^{\circ}$ and 1.0$^{\circ}$). The spectral index of -2.0 is chosen both because it is the standard index used in HAWC for studying extended sources~\cite{Abeysekara2017b} and because it is an expected index for PeVatrons.

A bright source may be found in the catalog search up to six times~\cite{Abeysekara2017b} (the three morphologies times two energy thresholds). To obtain one definitive source location and extension, the Right Ascension, Declination, and extension are simultaneously fit for each source in the $>$ 56 TeV map under the assumption of an E$^{-2.0}$ spectrum. These results are insensitive to the spectral index. A Gaussian spatial morphology is assumed. Because this is the first HAWC catalog constructed using maps with a high-energy threshold, we use the prefix ``eHWC'' (energy-HAWC) to identify the sources.

The bins above 56 TeV are then fit to a power-law shape with the spectral index fixed to -2.7. The extent is fixed to the fitted high-energy extent. This index typically gives a higher TS value, possibly indicating a steepening of the spectra at the highest energies. The integral flux above 56 TeV is computed using the result of this fit. For sources that are significantly detected above an estimated energy of 100 TeV, spectral fits to the emission over the whole energy range accessible to HAWC are also performed using a binned-likelihood forward-folding technique that takes into account the angular response of the detector as well as the bias and energy resolution of the energy estimator. 

When fitting the emission spectra of the sources, we do not consider multi-source or multi-component models; instead we fit the spectrum in the region of interest (3$^{\circ}$ radius) while assuming Gaussian-shaped emission and allowing the  value of the width to float. Contributions from diffuse emission and/or unresolved sources are not separated out. This introduces a systematic in the spectrum~\cite{Abeysekara2017b}. The integral flux values above 56 TeV are not expected to be affected since the diffuse emission falls rapidly with energy. In many cases, there are known to be two or more components to the emission, which may also affect the reported values of integral fluxes. For example, the eHWC J2030+412 region has contributions from both a pulsar wind nebula (PWN) and the possible TeV counterpart of the Fermi cocoon~\cite{Hona2019}. 

\section{Results}

\begin{figure*}
\centering
\includegraphics[width=0.99\textwidth]{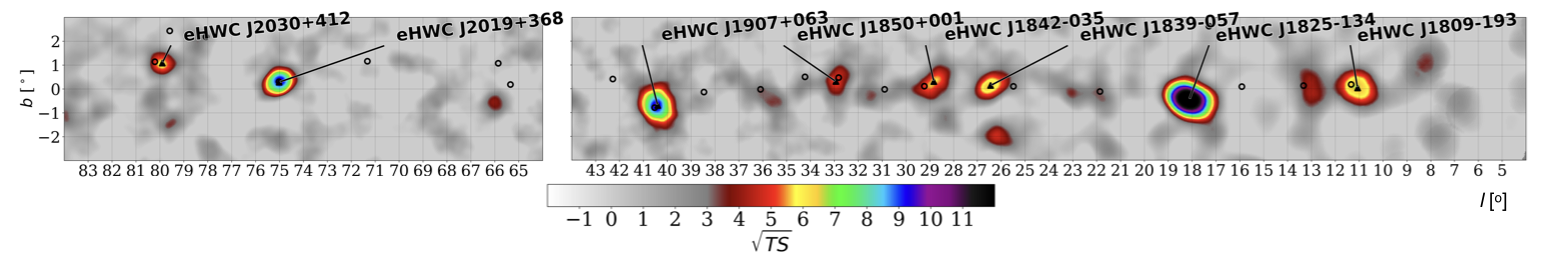}
\caption{$\sqrt{TS}$ map of the Galactic plane for $\hat{E} > 56$ TeV emission. A disk of radius 0.5$^{\circ}$ is assumed as the morphology. Black triangles denote the high-energy sources. For comparison, black open circles show sources from the 2HWC catalog. }
\label{fig:gr56}
\end{figure*}

\begin{figure*}
\centering
\includegraphics[width=0.99\textwidth]{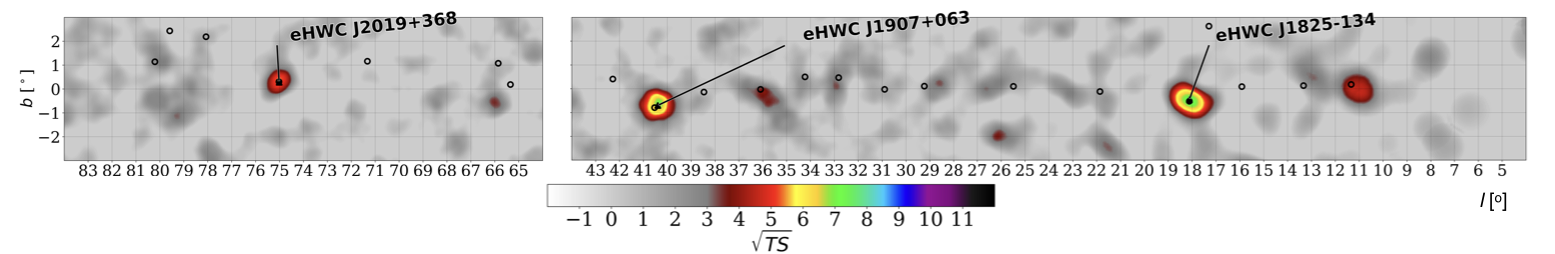}
\caption{The same as Figure \ref{fig:gr56}, but for $\hat{E} > 100$ TeV. The symbol convention is identical to Figure \ref{fig:gr56}.}
\label{fig:gr100}
\end{figure*}

\begin{table*}
\renewcommand{\arraystretch}{1.15}
\begin{center}
 \begin{tabular}{ | c || c| c | c | c| c | c | c | c| }
 \hline
 Source name & RA ($^o$) & Dec ($^o$) & Extension $>$ & F (10$^{-14}$ &  $\sqrt{TS}$ $>$ & nearest 2HWC & Distance to & $\sqrt{TS}$ $>$\\ 
  & & & 56 TeV ($^{o}$) & ph cm$^{-2}$ s$^{-1}$) & 56 TeV & source & 2HWC source($^{\circ}$) & 100 TeV \\
  \hline
   eHWC J0534+220 & 83.61 $\pm$ 0.02 & 22.00 $\pm$ 0.03 & PS & 1.2 $\pm$ 0.2 & 12.0 & J0534+220 & 0.02 & 4.44 \\ 
  
   eHWC J1809-193 & 272.46 $\pm$ 0.13 & -19.34 $\pm$ 0.14 & 0.34 $\pm$ 0.13 & 2.4$^{+0.6}_{-0.5}$ & 6.97  & J1809-190 & 0.30 & 4.82 \\

  eHWC J1825-134 & 276.40 $\pm$ 0.06 & -13.37 $\pm$ 0.06 & 0.36 $\pm$ 0.05 & 4.6 $\pm$ 0.5 & 14.5 & J1825-134 & 0.07 & 7.33 \\

  eHWC J1839-057 & 279.77 $\pm$ 0.12 & -5.71 $\pm$ 0.10 & 0.34 $\pm$ 0.08 & 1.5 $\pm$ 0.3 & 7.03 & J1837-065 & 0.96 & 3.06 \\
 
  eHWC J1842-035 & 280.72 $\pm$ 0.15 & -3.51 $\pm$ 0.11 & 0.39 $\pm$ 0.09 & 1.5 $\pm$ 0.3 & 6.63 & J1844-032 & 0.44 & 2.70 \\
 
  eHWC J1850+001 & 282.59 $\pm$ 0.21 & 0.14 $\pm$ 0.12 & 0.37 $\pm$ 0.16 & 1.1$^{+0.3}_{-0.2}$ & 5.31 & J1849+001 & 0.20 & 3.04 \\
    
  eHWC J1907+063 & 286.91 $\pm$ 0.10 & 6.32 $\pm$ 0.09 & 0.52 $\pm$ 0.09 & 2.8 $\pm$ 0.4 & 10.4 & J1908+063 & 0.16 & 7.30 \\ 
 
  eHWC J2019+368 & 304.95 $\pm$ 0.07 & 36.78 $\pm$ 0.04 & 0.20 $\pm$ 0.05 & 1.6$^{+0.3}_{-0.2}$ & 10.2 & J2019+367 & 0.02 & 4.85 \\
 
  eHWC J2030+412 & 307.74 $\pm$ 0.09 & 41.23 $\pm$ 0.07 & 0.18 $\pm$ 0.06 & 0.9 $\pm$ 0.2 & 6.43 & J2031+415 & 0.34 & 3.07 \\
  \hline
\end{tabular}
\caption{Sources exhibiting $\hat{E} >$ 56 TeV emission. A Gaussian morphology is assumed for a simultaneous fit to the source location and extension (68$\%$ Gaussian containment) for $\hat{E} >$ 56 TeV. The integral flux $F$ above 56 TeV is then fitted; $\sqrt{TS}$ is the square root of the test statistic for the integral flux fit. The nearest source from the 2HWC catalog and the angular distance to it are also provided. In addition, the $\sqrt{TS}$ of the same integral flux fit but above $\hat{E} >$100\,TeV is provided. All uncertainties are statistical only. The point spread function of HAWC for $\hat{E} >$ 56 TeV is $\sim$0.2$^{\circ}$ at the Crab declination~\cite{crab2018}, but is declination-dependent and increases to 0.35$^{\circ}$ and 0.45$^{\circ}$ for eHWC J1825-134 and eHWC J1809-193
respectively. The overall pointing error is 0.1$^{\circ}$~\cite{Abeysekara2017b}.}
\end{center}
\end{table*}

There are nine sources detected in the catalog search with significant ($\sqrt{TS}$ $>$ 5) emission for $\hat{E} >$ 56 TeV (see Table S1 of Supplemental Material~\cite{supp} for the results of the search). Eight of these sources are within $\sim$1$^{\circ}$ of the Galactic plane and are extended in apparent size (larger than HAWC's PSF) above this energy threshold. The only point source is the Crab Nebula (eHWC J0534+220), discussed in depth in \cite{crab2018}. Three of the sources show significant emission continuing above 100 TeV.

Figures \ref{fig:gr56} and \ref{fig:gr100} show $\sqrt{TS}$ maps of the Galactic plane for $\hat{E} >$ 56 TeV and $>$ 100 TeV, respectively. For the Crab Nebula, see Figure S1 in Supplemental Material~\cite{supp}. The sources are modeled as disks of radius 0.5$^{\circ}$. \mbox{Table I} gives the integral flux for $\hat{E} >$ 56 TeV for each source along with the fitted coordinates and Gaussian extension.

Most sources are within 0.5$^{\circ}$ of sources from the 2HWC catalog and, since they are extended, have overlapping emission.  We previously estimated a false positive rate of 0.5 all-sky sources~\cite{Abeysekara2017b}. However, all of the sources discussed here are located close to the Galactic plane and are consistent with previously known bright TeV sources, which makes them more likely to be the continuation of emission from lower energies than fluctuations.

Eight of the ten brightest sources from the 2HWC catalog are observed here. It is possible that ultra-high-energy emission is a generic feature of astrophysical sources and more sources will be discovered as more data are collected and more sensitive experiments are built. This raises questions about emission mechanisms of astrophysical sources, especially if they are leptonic in origin (see Discussion).

\begin{figure}
\centering
\includegraphics[width=0.48\textwidth]{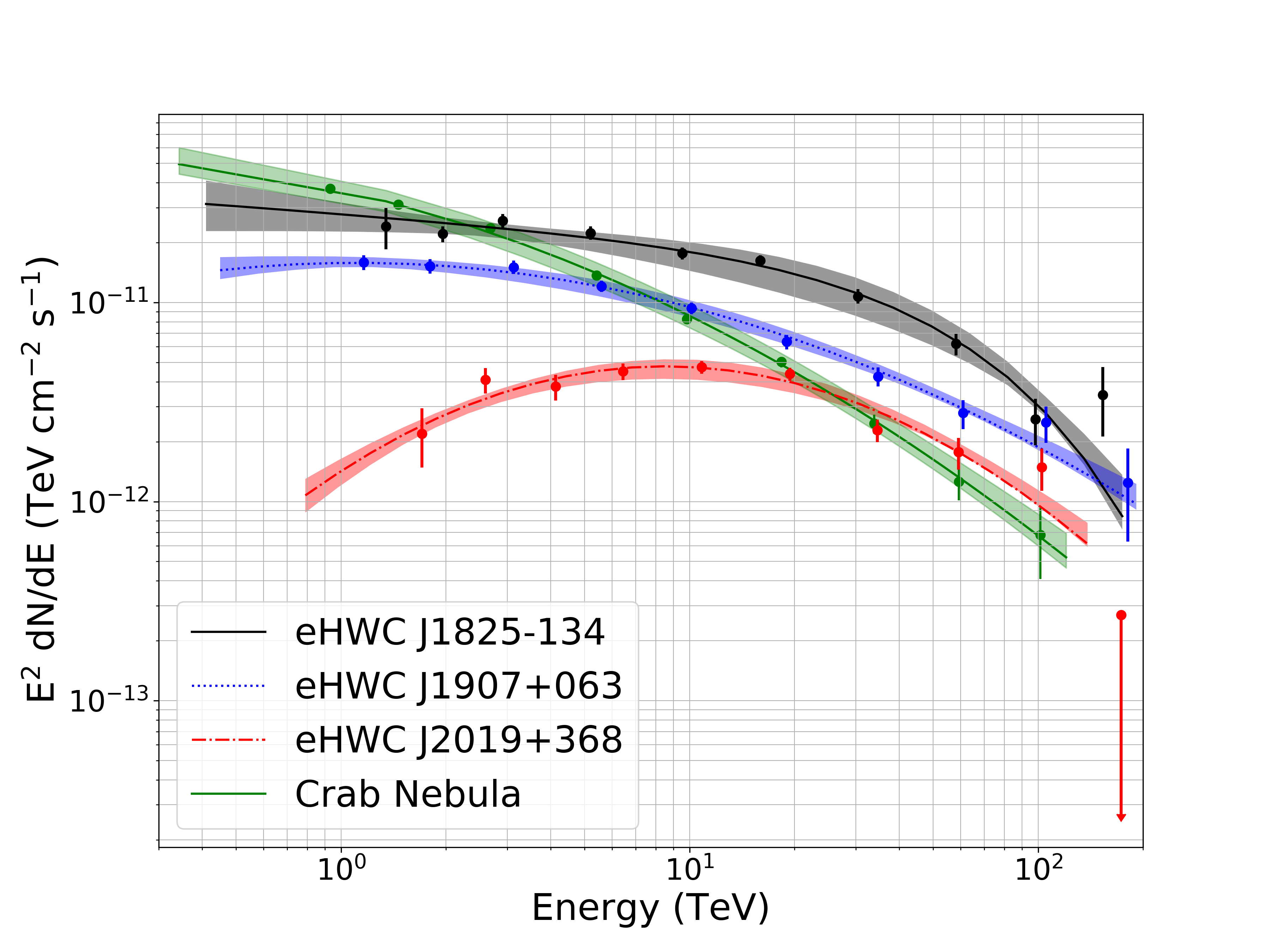}
\caption{The spectra of the three sources exhibiting significant $\hat{E} >$ 100 TeV emission. For each source, the line is the overall forward-folded best fit. The error bars on the flux points are statistical uncertainties only. The shaded band around the overall best fit line shows the systematic uncertainties related to the HAWC detector model, as discussed in \cite{crab2018}. The Crab Nebula spectrum from \cite{crab2018} is shown for comparison. }
\label{fig:spectra}
\end{figure}

Each source showing significant emission for $\hat{E} >$ 100 TeV is fit to three different spectral models: a power-law, a power-law with an exponential cutoff, and a log-parabola. For eHWC J1825-134, the most-probable model (using the Bayesian information criterion~\cite{gideon1978}) is a power-law with an exponential cutoff:
\begin{equation}\label{eq:plc}
\frac{dN}{dE} = \phi_{0} \left(\frac{E}{\mbox{10 TeV}}\right)^{-\alpha} \mathrm{exp}(-E/E_{cut}),
\end{equation}
while eHWC J1907+063 and eHWC J2019+368 are better fit to log-parabolas:
\begin{equation}\label{eq:lp}
\frac{dN}{dE} = \phi_{0} \left(\frac{E}{\mbox{10 TeV}}\right)^{-\alpha - \beta \mathrm{ln}(E/\mbox{10 TeV})},
\end{equation}

All three sources are extended in apparent size over HAWC's entire energy range.  Flux points are calculated for quarter-decade energy bins using the method described in \cite{crab2018}. When fitting the differential flux, it is assumed that the size of the source does not change with energy. Table II shows best-fit parameter values for these sources; Figure \ref{fig:spectra} shows their spectra.

\begin{table*}
\renewcommand{\arraystretch}{1.15}
\begin{center}
 \begin{tabular}{| m{2.6cm} || m{1.1cm} | m{2.6cm} |m{4cm} | m{2.6cm} | m{2cm} | m{1cm} |} 
 \hline
 Source & $\sqrt{TS}$ & Extension ($^{o}$) & $\phi_{0} $ ($10^{-13} $ TeV cm$^2$ s)$^{-1}$ & $\alpha$ & $E_{cut}$ (TeV) & PL diff\\ [0.5ex] 
  \hline
 eHWC J1825-134 & 41.1 & 0.53 $\pm$ 0.02 & 2.12 $\pm$ 0.15 & 2.12 $\pm$ 0.06 & 61 $\pm$ 12 &  7.4   \\
\hline\hline
 Source & $\sqrt{TS}$ & Extension ($^{o}$) & $\phi_{0}$ (10$^{-13}$ TeV cm$^2$ s)$^{-1}$ & $\alpha$ & $\beta$ & PL diff\\ [0.5ex] 
 \hline
 eHWC J1907+063 & 37.8 & 0.67 $\pm$ 0.03 & 0.95 $\pm$ 0.05 & 2.46 $\pm$ 0.03 & 0.11 $\pm$ 0.02 & 6.0 \\

 eHWC J2019+368 & 32.2 & 0.30 $\pm$ 0.02 & 0.45 $\pm$ 0.03 & 2.08 $\pm$ 0.06 & 0.26 $\pm$ 0.05 & 8.2 \\
 \hline
\end{tabular}
 \caption{Spectral fit values for the three sources that emit above 100 TeV. eHWC J1825-134 is fit to a power-law with an exponential cutoff (Eq. \ref{eq:plc}); the other two sources are fit to a log-parabola (Eq. \ref{eq:lp}). $\sqrt{TS}$ is the square root of test statistic for the given likelihood spectral fit. Sources are modeled as a Gaussian; \textit{Extension} is the Gaussian width over the entire energy range. The uncertainties are statistical only. $\phi_0$ is the flux normalization at the pivot energy (10 TeV). \textit{PL diff} gives $\sqrt{\Delta\mbox{TS}}$ between the given spectral model and a power-law.   }
\end{center}
\end{table*}

We investigated whether the observed high-energy detections are compatible with being entirely due to mis-reconstructed events (see Tables S3 and S4 in Supplemental Material~\cite{supp}). For eHWC J1907+063, the strongest highest-energy detection, emission above a true energy of 56 TeV (100 TeV) is detected at the 7.6$\sigma$ (4.6$\sigma$) level. Note that this is more conservative than the procedure followed in \cite{Amenomori2019}.

Each of the three $>$ 100 TeV regions have also been observed by at least one of the imaging atmospheric Cherenkov telescopes (IACTs) (References: eHWC J1825-134~\cite{Abdalla2018,Anguner2017},  eHWC J1907+063~\cite{Aliu2014, Aharonian2009}, eHWC J2019+368~\cite{Aliu2014b,TheVERITASCollaboration2018}). The HAWC measurements extend the energy range of these sources past 100 TeV for the first time. HAWC tends to measure higher fluxes ($\sim$2x difference) and larger source extents than the IACT measurements. These discrepancies cannot be explained by a misunderstanding of the HAWC detector response, as the HAWC spectrum of the Crab Nebula agrees with IACT measurements within uncertainties~\cite{crab2018}. 

Both eHWC J2019+368 and eHWC J1825-134 have been separated into two or more sources by IACTs (see Table S8 in Supplemental Material~\cite{supp} for a list of TeVCat sources within 3$^{\circ}$ of each source), and the HAWC emission is the sum of these plus any additional unresolved sources. For example, eHWC J1825-134 overlaps with both HESS J1825-137 and HESS J1826-130. There are also differences in the computation of the background estimate~\cite{Abeysekara2018,Jardin-blicq} as well as the fact that contributions from diffuse emission are not considered here. This will be addressed in future papers.

\section{Discussion}
Although Klein-Nishina effects mean that any IC component of the emission becomes suppressed beginning around 10 TeV, merely detecting high-energy emission is not enough to claim a hadronic emission origin. The Crab Nebula is a firmly-identified electron accelerator~\cite{DeJager1996} that emits well past 100 TeV~\cite{Amenomori2019,crab2018}. We consider both hadronic and leptonic emission mechanisms.

\subsection{Leptonic emission mechanisms}

All nine sources have at least one pulsar from the Australia Telsecope National Facility (ATNF) database~\cite{Manchester2005} within 0.5 degrees of the HAWC high-energy location (see \mbox{Table III}). Borrowing the terminology coined in \cite{Linden2017,Linden2018}, it has been suggested that these gamma-ray sources may be ``TeV halos''. The spatial extents of these objects are much larger than the X-ray PWN ($\sim$25 pc) and the emission is leptonic in origin, stemming from electrons that have escaped the PWN radius~\cite{Sudoh2019}. For eight of these nine sources, at least one nearby pulsar has an extremely high spin-down power ($\dot{E}$ $> 10^{36}$ erg/s). The distance between the center of the HAWC high-energy emission and the pulsar is generally less than the extent of the HAWC source.

\begin{table*}
\renewcommand{\arraystretch}{1.15}
\begin{center}
 \begin{tabular}{ | c || c| c | c | c| c | c | }
 \hline
 HAWC source & PSR name & $\dot{E}$ & Age ($\frac{P}{2\dot{P}}$) & Distance to  & Distance between HAWC  & HAWC source \\  
 & & (erg/s) &(kyr) &  Earth (kpc) & source and PSR [$^{\circ}$ (pc)] & extent (pc) \\
  \hline
eHWC J0534+220 & J0534+2200 & 4.5$\times$10$^{38}$ & 1.3 & 2.00 & 0.03 (1.05) & - \\
 
  eHWC J1809-193 & J1809-1917 & 1.8$\times$10$^{36}$ & 51.3 & 3.27 & 0.05 (2.86) & 19.4 \\ 
  - & J1811-1925 & 6.4$\times$10$^{36}$ & 23.3 & 5.00 & 0.40 (34.9) & 29.7 \\ 
 
  eHWC J1825-134 & J1826-1334 & 2.8$\times$10$^{36}$ & 21.4 & 3.61 & 0.26 (16.4) & 22.1 \\
  - & J1826-1256 & 3.6$\times$10$^{36}$  & 14.4 & 1.55 & 0.45 (12.2) & 9.47 \\

  eHWC J1839-057 & J1838-0537 & 6.0$\times$10$^{36}$ & 4.89 & 2.0\footnote{Pseudo-distance from \cite{Pletsch2012} } & 0.10 (3.50) & 11.9 \\
 
  eHWC J1842-035 & J1844-0346 & 4.2$\times$10$^{36}$ & 11.6 & 2.40\footnote{Pseudo-distance from Eq. 3 of \cite{Wu2018}} & 0.49 (20.5) & 16.3 \\
 
  eHWC J1850+001 & J1849-0001 & 9.8$\times$10$^{36}$ & 42.9 & 7.00\footnote{Distance estimate from \cite{Gotthelf2011}}  & 0.37 (45.2) & 45.2 \\

  eHWC J1907+063 & J1907+0602 & 2.8$\times$10$^{36}$ & 19.5 & 2.37 & 0.29 (12.0) & 21.5 \\

  eHWC J2019+368 & J2021+3651 & 3.4$\times$10$^{36}$ & 17.2 & 1.80 & 0.27 (8.48) & 6.28 \\
 
  eHWC J2030+412 & J2032+4127 & 1.5$\times$10$^{35}$ & 201 & 1.33 & 0.33 (7.66) & 4.18 \\
  \hline
\end{tabular}
\caption{Information on all pulsars with $\dot{E} > 10^{36}$ erg/s within 0.5 degree of each source. The only pulsar within 0.5 degree of eHWC J2030+412 has an $\dot{E}$ below this threshold; it is included here for completeness. All pulsar parameters come from the ATNF database, version 1.60~\cite{Manchester2005} unless specified. The distance between the pulsar and the HAWC source as well as the HAWC high-energy source extent (from Table I) are given in parsecs here, assuming that the HAWC source is the same distance from the Earth as the pulsar. } 
\end{center}
\end{table*}

There are only 26 high-$\dot{E}$ pulsars in the inner Galactic plane ($|b| <$ 1$^{\circ}$ in Galactic coordinates) and within HAWC's field-of-view (roughly $0^{\circ} < l < 90^{\circ}$). Depending on the spatial distribution of pulsars assumed, we expect only $\sim$1-2 pulsars to be within 0.5$^{\circ}$ of a HAWC high-energy source by chance. The Crab is not located in the inner Galactic plane and is therefore excluded from this calculation, but is also associated with a high-$\dot{E}$ pulsar.

If these sources are all leptonic in nature, their extension raises interesting questions about particle diffusion as the highest-energy electrons are expected to cool very quickly, before traveling large distances.

The electrons that produce the gamma rays will also radiate synchrotron emission in X-rays. To produce 100 TeV gamma rays on the cosmic microwave background requires electrons of $\sim$300 TeV, resulting in synchrotron emission peaking at 10 keV in a magnetic field of 3 microgauss~\cite{Hinton2009}. Dedicated analyses including multi-wavelength studies will be part of upcoming publications on individual objects.

\subsection{Hadronic emission mechanisms} 

Hadronic emission mechanisms could also contribute, even if the emission is dominantly leptonic. Assuming that these sources are connected to the pulsars, they are all fairly young, with the mean (median) characteristic age being 37 (20) kyr. This means that the observed TeV emission may include a contribution from a supernova remnant~\cite{Linden2017}. 

All three source spectra presented here either have a cutoff or curvature before 100 TeV, preventing their unambiguous identification as PeVatrons. However, this does not immediately disfavor the PeVatron hypothesis, since spectral curvature might already be present at tens of TeV~\cite{Aharonian2013} and additional steepening of the high-energy tails may be expected from pair production on the interstellar radiation field and the cosmic microwave background~\cite{Porter2018}. Additionally, the reported spectra here may include contributions from multiple sources, which makes it harder to interpret the cutoff as it relates to the nature of the gamma-ray emission.

If the emission is due to hadronic mechanisms, these gamma-ray sources may be potential neutrino sources~\cite{Halzen2017}. Two sources are especially interesting:

An IceCube search for point-like sources in the astrophysical neutrino flux, the eHWC J1907+063 region had the second-best p-value (although still consistent with a background-only hypothesis) in an \textit{a priori} defined source list motivated by gamma-ray observations~\cite{Aartsen2019}. The HAWC spectrum presented here, which has a relatively hard spectral index and less curvature than the other sources, provides hints of a hadronic component.

The eHWC J2030+412 region is coincident with the Cygnus OB2 complex, which is one of the young massive star clusters that has been previously suggested as a site of CR acceleration~\cite{Aharonian}. 

\section{Conclusions}
We report HAWC observations of the highest-energy gamma-ray sources to date. There are nine sources with $\hat{E} >$ 56 TeV emission; three also have significant $\hat{E} >$ 100 TeV emission. Emission mechanisms are not yet clear, especially for eHWC J1825-134 and eHWC J1907+063. These are the two most significant sources above 100 TeV and both exhibit relatively hard spectra with extension at the highest energies, as expected for PeVatrons. Forthcoming HAWC observations of these sources ~\cite{Brisbois2019,Hona2019,Salesa2019} combined with multi-messenger and multi-wavelength studies will be important in disentangling emission mechanisms. 

\section{Acknowledgements}
\begin{acknowledgments}
We acknowledge the support from: the US National Science Foundation (NSF); the US Department of Energy Office of High-Energy Physics; 
the Laboratory Directed Research and Development (LDRD) program of Los Alamos National Laboratory; 
Consejo Nacional de Ciencia y Tecnolog\'{\i}a (CONACyT), M{\'e}xico, grants 271051, 232656, 260378, 179588, 254964, 258865, 243290, 132197, A1-S-46288, c{\'a}tedras 873, 1563, 341, 323, Red HAWC, M{\'e}xico; 
DGAPA-UNAM grants IG100317, IN111315, IN111716-3, IA102715, IN109916, IA102917, IN112218; 
VIEP-BUAP; 
PIFI 2012, 2013, PROFOCIE 2014, 2015; 
the University of Wisconsin Alumni Research Foundation; 
the Institute of Geophysics, Planetary Physics, and Signatures at Los Alamos National Laboratory; 
Polish Science Centre, grants DEC-2018/31/B/ST9/01069, DEC-2017/27/B/ST9/02272; 
Coordinaci{\'o}n de la Investigaci{\'o}n Cient\'{\i}fica de la Universidad Michoacana; Royal Society - Newton Advanced Fellowship 180385; 
Thanks to Scott Delay, Luciano D\'{\i}az and Eduardo Murrieta for technical support.

\end{acknowledgments}

\clearpage

\setcounter{equation}{0}
\setcounter{table}{0}
\setcounter{figure}{0}
\renewcommand{\theequation}{S\arabic{equation}}
\renewcommand{\thetable}{S\arabic{table}}
\renewcommand{\thefigure}{S\arabic{figure}}

\section{Supplemental Material for Multiple Galactic Sources with Emission Above 56 TeV Detected by HAWC}

\section{Test statistic}

Throughout this Letter, the test statistic is defined as twice the likelihood ratio:
\begin{equation}
TS = 2 \ln \left( \frac{L_{s+b}}{L_b} \right)\enspace,
\end{equation}
where $L_{s+b}$ is the best-fit likelihood for the signal plus background hypothesis, while $L_b$ is the null (background-only) hypothesis.  $L$ is the standard definition of likelihood from~\cite{Younk2015}.

When Wilks' theorem~\cite{Wilks1938} is assumed (a valid assumption for HAWC data), this TS is distributed as a chi-square with the number of degrees of freedom equal to the difference in the number of free parameters between the hypothesis. When the number of free parameters is unity, $\sqrt{TS}$ can be interpreted as Gaussian significance~\cite{Younk2015}. Such is the case in the catalog search presented here, as the only free parameter in the maps shown in Figures 1 and 2 is the flux normalization. 

\section{Catalog search results}
Table S1 lists each source found in the high-energy catalog search, along with the search in which it was found (point source or extended), its coordinates, the closest 2HWC source, and the angular distance between those coordinates and the 2HWC source location. 

\begin{table*}
\begin{center}
 \begin{tabular}{| c | c | c | c |c| c |c|c |} 
 \hline
 Source & Energy & Radius & $\sqrt{TS}$ & RA ($^{\circ}$) & Dec ($^{\circ}$) & Nearest 2HWC source & Distance to 2HWC source ($^{\circ}$)\\
 \hline
 eHWC J0534+220 & $>$ 56 TeV & PS & 11.6 & 83.63 & 22.02 & 2HWC 0534+220 (Crab) & 0.0 \\
 eHWC J0534+220 & - & 0.5$^{\circ}$ & 9.42 & 83.67 & 22.06 & - & 0.06 \\
 eHWC J0535+220 & - & 1.0$^{\circ}$ & 5.72 & 83.94 & 22.06 & - & 0.31 \\
 \hline
  eHWC J1809-193 & $>$ 56 TeV & PS & 6.75 & 272.42 & -19.39 & 2HWC J1809-190 & 0.35 \\
  eHWC J1809-193 & - & 0.5$^{\circ}$ & 6.73 & 272.46 & -19.31 & - & 0.27 \\
  eHWC J1810-192 & - & 1.0$^{\circ}$ & 6.30 & 272.55 & -19.27 & - & 0.25 \\
  \hline 
  eHWC J1825-132 & $>$ 56 TeV & PS & 12.1 & 276.46 & -13.25 & 2HWC J1825-134 & 0.15 \\
  eHWC J1825-133 & - & 0.5$^{\circ}$ & 13.9 & 276.42 & -13.36 & - & 0.06 \\
  eHWC J1824-133 & - & 1.0$^{\circ}$ & 13.2 & 276.24 & -13.36 & - & 0.22 \\
  eHWC J1825-132 & $>$ 100 TeV & PS & 6.75 & 276.42 & -13.21 & - & 0.19  \\
  eHWC J1825-134 & - & 0.5$^{\circ}$ & 7.34 & 276.42 & -13.44 & - & 0.06 \\
  eHWC J1824-134 & - & 1.0$^{\circ}$ & 6.76 & 276.20 & -13.40 & - & 0.26 \\
  \hline 
  eHWC J1839-057 & $>$ 56 TeV & PS & 5.57 & 279.84 & -5.79 & 2HWC J1837-065 & 0.92 \\
  eHWC J1839-056 & - & 0.5$^{\circ}$ & 6.41 & 279.84 & -5.64 & - & 1.06  \\
  eHWC J1837-056 & - & 1.0$^{\circ}$ & 5.99 & 279.45 & -5.60 & - & 0.98  \\
  \hline 
  eHWC J1842-035 & $>$ 56 TeV & 0.5$^{\circ}$ & 5.74 & 280.68 & -3.51 & 2HWC J1844-032 & 0.47 \\
  eHWC J1843-036 & - & 1.0$^{\circ}$ & 5.85 & 280.85 & -3.66 & - & 0.47 \\
  \hline
  eHWC J1849+000 & $>$ 56 TeV & PS & 5.27 & 282.31 & 0.04 & 2HWC J1849+001 & 0.11 \\
  eHWC J1850+001 & - & 0.5$^{\circ}$ & 5.05 & 282.57 & 0.19 & - & 0.20 \\
  eHWC J1849-004 & - & 1.0$^{\circ}$ & 5.50 & 282.44 & -0.45 & - & 0.56 \\
  \hline
  eHWC J1908+065 & $>$ 56 TeV & PS & 7.27 & 287.01 & 6.50 & 2HWC J1908+063 & 0.12 \\
  eHWC J1907+063 & - & 0.5$^{\circ}$ & 9.54 & 286.96 & 6.39 & - & 0.09  \\
  eHWC J1907+062 & - & 1.0$^{\circ}$ & 9.63 & 286.83 & 6.20 & - & 0.29 \\
  eHWC J1908+065 & $>$ 100 TeV & PS & 5.70 & 287.01 & 6.50 & - & 0.12 \\
  eHWC J1907+063 & - & 0.5$^{\circ}$ & 7.03 & 286.79 & 6.32 & - & 0.27 \\
  eHWC J1908+065 & - & 1.0$^{\circ}$ & 6.71 & 287.01 & 6.50 & - &  0.12 \\
  \hline
  eHWC J2020+367 & $>$ 56 TeV & PS & 9.73 & 305.02 & 36.80 & 2HWC J2019+367 & 0.08 \\ 
  eHWC J2019+367 & - & 0.5$^{\circ}$ & 9.48 & 304.85 & 36.80 & - & 0.08 \\
  eHWC J2019+371 & - & 1.0$^{\circ}$ & 7.15 & 304.89 & 37.12 & - & 0.32 \\
  eHWC J2019+367 & $>$ 100 TeV & PS & 5.39 & 304.94 & 36.80 & - & 0.00 \\
  eHWC J2019+368 & - & 0.5$^{\circ}$ & 5.36 & 304.89 & 36.84 & - & 0.06 \\
  \hline
  eHWC J2030+412 & $>$ 56 TeV & PS & 5.74 & 307.70 & 41.26 & 2HWC J2031+415 & 0.34 \\
  eHWC J2031+412 & - & 0.5$^{\circ}$ & 5.94 & 307.84 & 41.21 & - & 0.31   \\
 \hline
 \end{tabular}
 \caption{The results of the high-energy catalog search. The column ``Radius" denotes which map the source was found in. PS stands for ``point source".  $\sqrt{TS}$ is the square root of the test statistic of detection, assuming a power law spectrum with an index of 2.0 and the specified source radius. The distance between each source and the nearest 2HWC source is also given. After the catalog search is run, each source's Right Ascension, Declination, and extension are concurrently fit; these results are in Table I of the main text.}
 \end{center}
 \end{table*}
 
\section{Crab Nebula figures}

\begin{figure*}
\centering
\includegraphics[width=0.48\textwidth]{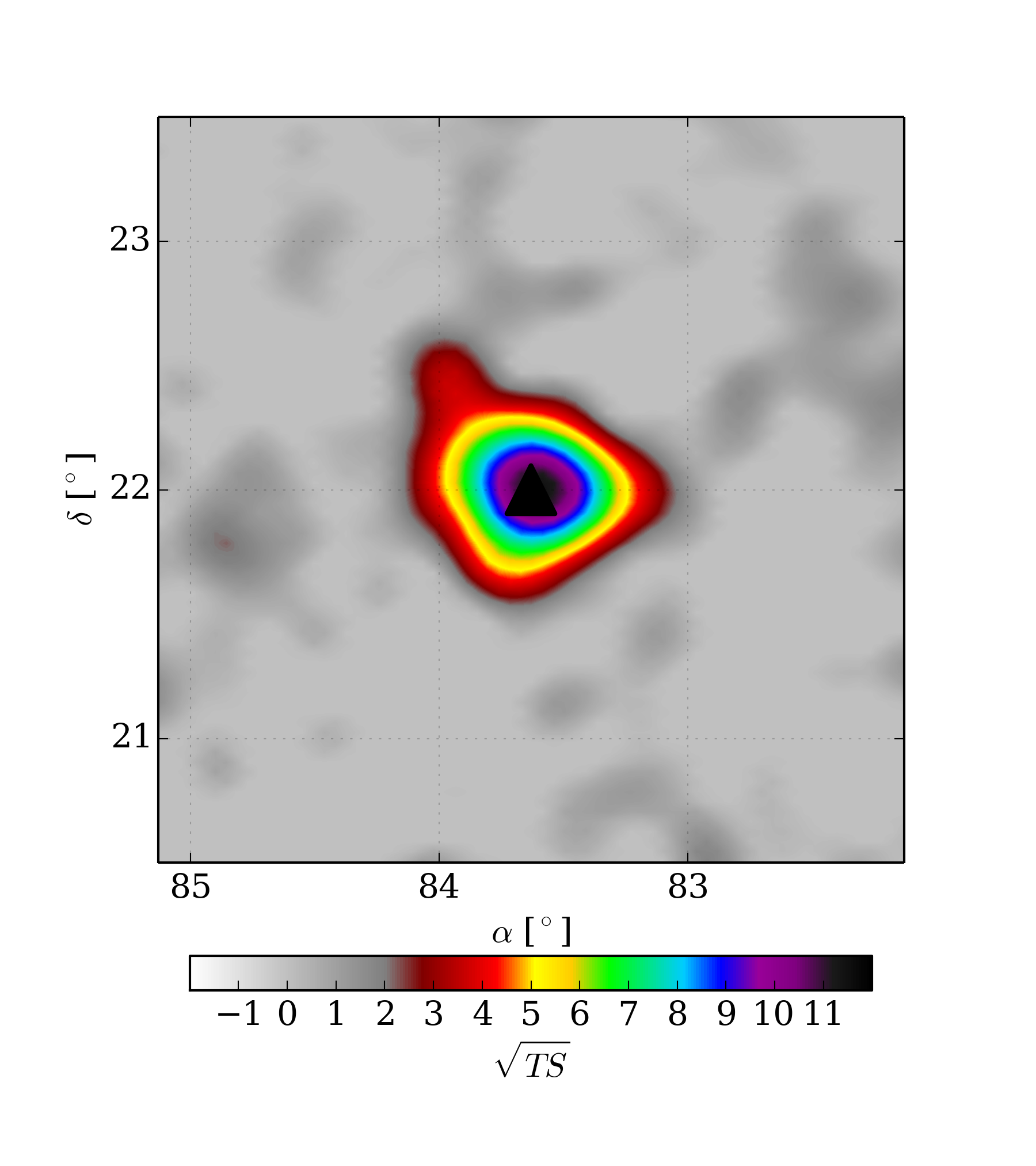}
\subfloat{\includegraphics[width=0.48\textwidth]{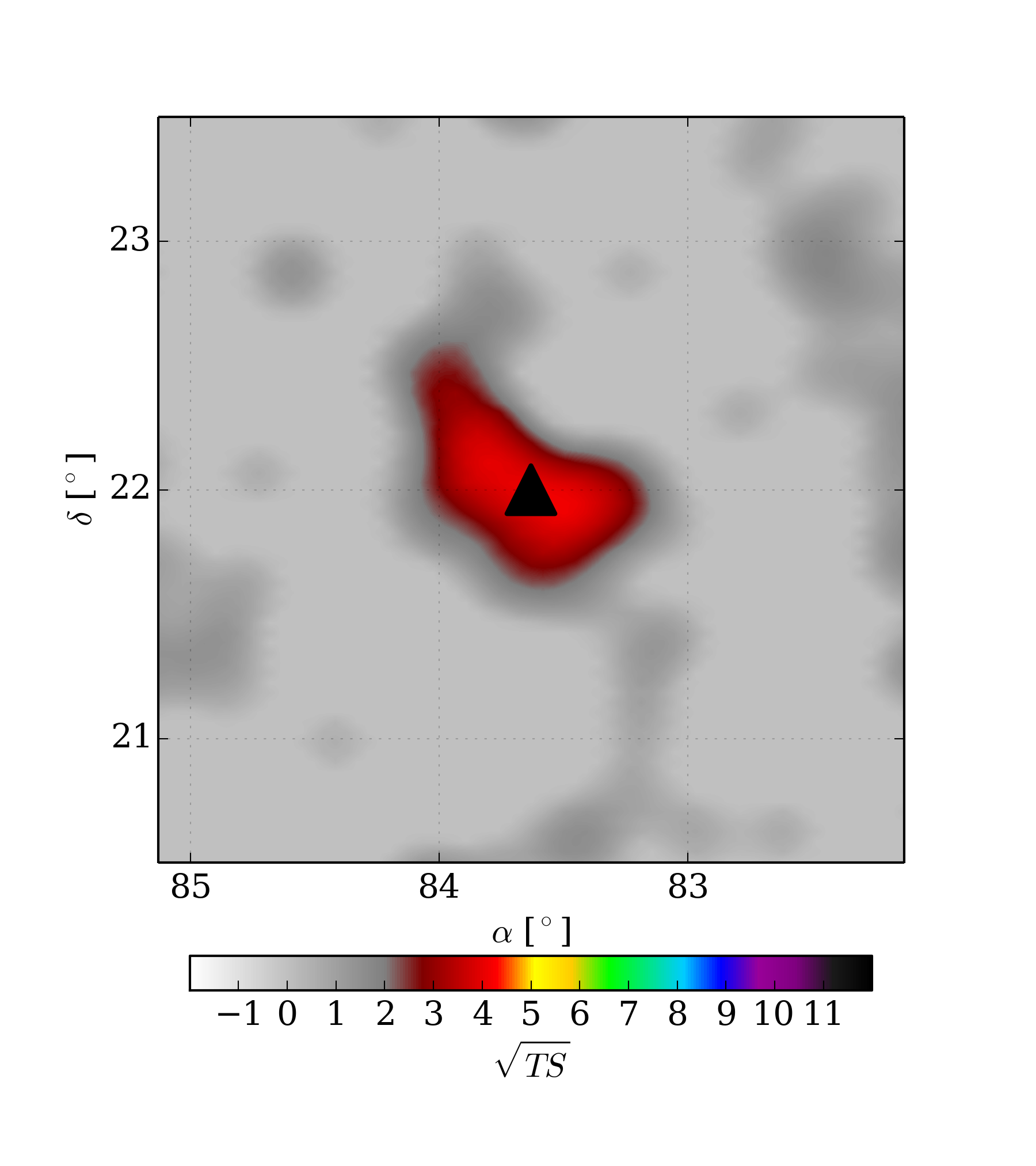}}
\caption{The $\sqrt{TS}$ map of the region around the Crab Nebula for $\hat{E} >$ 56 TeV (left) and $>$ 100 TeV (right). The black triangle denotes the fitted location of the source above 56 TeV.} 
\label{fig:crab}
\end{figure*}

The $\sqrt{TS}$ map of the Crab Nebula above 56 TeV and 100 TeV can be seen in Figure \ref{fig:crab}.
 
 \section{Integral flux}
 
 The integral flux values in Table I are calculated assuming a power law with an index of -2.7. This index typically gives a higher TS value than harder indices, possibly indicating curvature or cutoffs in the spectra at the highest energies. When a spectral index of -2.0 is instead assumed, the average change in the integral flux value is $\sim$20$\%$. Table S2 gives these integral flux values calculated assuming a power law with an index of -2.0.
 
 \begin{table*}
\renewcommand{\arraystretch}{1.15}
\begin{center}
 \begin{tabular}{| c | c | c |c |} 
 \hline 
 Source & F (10$^{-14}$ ph cm$^{-2}$ s$^{-1}$) &  $\sqrt{TS}$ $>$ 56 TeV & $\sqrt{TS}$ $>$ 100 TeV\\
 \hline
 eHWC J0534+220 & 1.5$^{+0.3}_{-0.2}$ & 11.7 & 4.27 \\ 
 eHWC J1809-193 & 2.6$^{+0.7}_{-0.6}$ & 6.76 & 4.69 \\   
 eHWC J1825-134 & 5.4$^{+0.7}_{-0.6}$ & 14.0 & 7.35 \\   
 eHWC J1839-057 & 1.7$^{+0.4}_{-0.3}$ & 6.63 & 3.03 \\    
 eHWC J1842-035 & 1.7$^{+0.4}_{-0.3}$ & 6.06 & 2.52 \\    
 eHWC J1850+001 & 1.3$^{+0.4}_{-0.3}$ & 5.18 & 3.09 \\       
 eHWC J1907+063 & 3.4$^{+0.5}_{-0.4}$  & 10.3 & 7.17 \\    
 eHWC J2019+368 & 1.9 $\pm$ 0.3 & 9.86 & 5.55 \\    
 eHWC J2030+412 & 1.1$^{+0.3}_{-0.2}$ & 6.16 & 2.96 \\
 \hline 
 \end{tabular}
 \caption{The integral flux values calculated assuming a spectral index of -2.0. An index of -2.7 is assumed in the main text.}
 \end{center}
 \end{table*}
 
 \section{Hard-cutoff fits}
 Not all of the events in an estimated energy bin have true energies in that bin. Some of these events may be mis-reconstructed low energy events that have migrated into a higher-energy bin. Since astrophysical sources emit following roughly power law spectra, in some cases there may be more mis-reconstructed low energy events than true high-energy events.
 
To ensure that these sources are true $>$ 56 TeV detections and that we have not simply detected mis-reconstructed lower energy gamma-rays, the integral flux above 56 TeV calculation was repeated for a spectral shape of a power-law convolved with a hard cutoff (step function) at 56 TeV. For the sources emitting significantly above 100 TeV in reconstructed energy, this process was repeated with the hard cutoff moved to 100 TeV. The TS for integral flux fit (also given in Table I of the main text), as well as this version with the hard cutoff, are given in Table S3. In all cases, the fit without the hard cutoff is preferred; the strongest detection is eHWC J1907+063.

For the three sources where spectra are computed (eHWC J1825-134, eHWC J1907+063, and eHWC J2019+368), the hard-cutoff fit is also convolved with the best-fit spectral model over HAWC's entire energy range (beginning at $\hat{E}$ = 1 TeV). These results are given in Table S4. Once again, eHWC J1907+063 is the most significant highest-energy source detection.

\begin{table*}
\renewcommand{\arraystretch}{1.15}
\begin{center}
 \begin{tabular}{| c | c | c | c |c| c |c|c |} 
 \hline
 Source & TS & TS$_{56}$ & $\sqrt{\Delta TS_{56}}$ & TS$_{100}$ & $\sqrt{\Delta TS_{100}}$\\
 \hline
 eHWC J0534+220 & 143.7 & 130.1 & 3.7 & - & -  \\ 

 eHWC J1809-193 & 48.6 & 31.4 & 4.1 & - & - \\

 eHWC J1825-134 & 210.8 & 177.5 & 5.8 & 198.0 & 3.4 \\ 

 eHWC J1839-057 & 49.4 & 44.2 & 2.3 & - & - \\

 eHWC J1842-035 & 44.0 & 40.3 & 1.9 & - & - \\

 eHWC J1850+001 & 28.2 & 20.3 & 2.8 & - & - \\

 eHWC J1907+063 & 108.7 & 63.0 & 6.8 & 88.7 & 4.5 \\

 eHWC J2019+368 & 104.6 & 77.7 & 5.2 & 102.3 & 1.5 \\

 eHWC J2030+412 & 41.4 & 33.7 & 2.8 & - & - \\ 
 \hline
 \end{tabular}
 \caption{TS values for three different fits. \textit{TS} is the test statistic for the calculation of the integral flux above 56 TeV, which assumes a power-law spectrum with an index of -2.7. \textit{TS$_{56}$} is the test statistic for that fit convolved with a step function at 56 TeV, while $\sqrt{\Delta TS_{56}}$ = $\sqrt{TS-TS_{56}}$. For the three sources detected above 100 TeV in reconstructed energy, this process is repeated for a step function inserted at 100 TeV (\textit{TS$_{100}$}). }
 \end{center}
 \end{table*}
 
\begin{table*}
\renewcommand{\arraystretch}{1.15}
\begin{center}
\begin{tabular}{| c | c | c | c |c| c |c|c |} 
\hline
Source & TS & TS$_{56}$ & $\sqrt{\Delta TS_{56}}$ & TS$_{100}$ & $\sqrt{\Delta TS_{100}}$\\ 
\hline 
eHWC J1825-134 & 1685.7 & 1643.2 & 6.5 & 1674.9 & 3.3 \\
eHWC J1907+063 & 1429.0 & 1370.0 & 7.6 & 1409.0 & 4.6 \\
eHWC J2019+368 & 1039.4 & 999.4 & 6.3  & 1032.5 & 2.6 \\
\hline
\end{tabular}
\caption{TS values for the full spectral fits for the three sources detected above 100 TeV. The table is identical to Table S3, except \textit{TS} is instead the test statistic for best-fit spectral shape (a power-law with an exponential cutoff for eHWC J1825-134 and a log-parabola for the other two sources). HAWC's entire energy range is taken into consideration here rather than just the energy bins above 56 TeV.}
\end{center}
\end{table*}

\section{Flux points for each high-energy source }
Tables S5 through S7 give the $\sqrt{TS}$, median energy, and flux in each reconstructed energy bin for the three sources where spectra are fit (eHWC J1825-134, eHWC J1907+063, and eHWC J2019+368). The binning scheme is defined in \cite{crab2018}. The quoted uncertainties on the flux points are statistical only. Upper limits (95$\%$ CL) are computed when the TS in an energy bin is less than 4.

Note that the median bin energy in a given $\hat{E}$ bin may fall outside the reconstructed energy bin ranges. This can happen for a variety of reasons and is not unexpected. The steepness of the spectrum affects the simulated energy to reconstructed energy correlation in each bin. Additionally, HAWC's energy resolution and energy bias are declination dependent.  \\

\begin{table*}
\renewcommand{\arraystretch}{1.15}
\begin{center}
 \begin{tabular}{| c | c | c |c| c |} 
 \hline
 Bin & $\hat{E}$ energy range (TeV) & $\sqrt{TS}$ & Median energy (TeV) & Flux (TeV cm$^{-2}$ s$^{-1}$) \\
 \hline
 c & 1.00-1.78 & 4.29 & 1.34 & (2.41 $\pm$ 0.56) $\times$ 10$^{-11}$ \\
 d & 1.78-3.16 & 11.3 &  1.96 & (2.22 $\pm$ 0.20) $\times$ 10$^{-11}$ \\
 e & 3.16-5.62 & 12.9 & 2.91 & (2.58 $\pm$ 0.21) $\times$ 10$^{-11}$ \\
 f & 5.62-10.0 & 14.7 & 5.20 & (2.24 $\pm$ 0.17) $\times$ 10$^{-11}$ \\
 g & 10.0-17.8 & 16.0 & 9.54 & (1.77 $\pm$ 0.13) $\times$ 10$^{-11}$ \\
 h & 17.8-31.6 & 20.2 & 15.95 & (1.63 $\pm$ 0.11) $\times$ 10$^{-11}$ \\
 i & 31.6-56.2 & 17.2 & 30.44 & (1.08 $\pm$ 0.09) $\times$ 10$^{-11}$ \\
 j & 56.2-100 & 12.6 & 58.18 & (6.22 $\pm$ 0.76) $\times$ 10$^{-12}$ \\
 k & 100-177 & 5.72 & 98.17 & (2.59 $\pm$ 0.68) $\times$ 10$^{-12}$ \\
 l & 177-316 & 4.03 & 153.5 & (3.45 $\pm$ 1.25) $\times$ 10$^{-12}$\\

 \hline
\end{tabular}
 \caption{Flux points for eHWC J1825-134. This source is fit to a power-law with an exponential cutoff. Uncertainties are statistical only.}
\end{center}
\end{table*}

\begin{table*}
\renewcommand{\arraystretch}{1.15}
\begin{center}
 \begin{tabular}{| c | c | c |c| c |} 
 \hline
 Bin & $\hat{E}$ energy range (TeV) & $\sqrt{TS}$ & Median energy (TeV) & Flux (TeV cm$^{-2}$ s$^{-1}$) \\
 \hline
 c & 1.00-1.78 & 11.7 & 1.16 & (1.59 $\pm$ 0.14) $\times$ 10$^{-11}$ \\
 d & 1.78-3.16 & 12.4 &  1.80 & (1.52 $\pm$ 0.13) $\times$ 10$^{-11}$ \\
 e & 3.16-5.62 & 13.7 & 3.13 & (1.51 $\pm$ 0.11) $\times$ 10$^{-11}$ \\
 f & 5.62-10.0 & 16.2 & 5.59 & (1.21 $\pm$ 0.08) $\times$ 10$^{-11}$ \\
 g & 10.0-17.8 & 16.3 & 10.13 & (9.36 $\pm$ 0.63) $\times$ 10$^{-12}$ \\
 h & 17.8-31.6 & 13.6 & 19.0 & (6.36 $\pm$ 0.53) $\times$ 10$^{-12}$ \\
 i & 31.6-56.2 & 11.4 & 34.79 & (4.25 $\pm$ 0.46) $\times$ 10$^{-12}$ \\
 j & 56.2-100 & 7.66 & 60.89 & (2.78 $\pm$ 0.46) $\times$ 10$^{-12}$ \\
 k & 100-177 & 6.54 & 105.4 & (2.49 $\pm$ 0.53) $\times$ 10$^{-12}$ \\
 l & 177-316 & 2.66 & 180.8 & (1.25 $\pm$ 0.61) $\times$ 10$^{-12}$\\

 \hline
\end{tabular}
 \caption{Flux points for eHWC J1907+063. This source is fit to a log parabola. Uncertainties are statistical only.}
\end{center}
\end{table*}

\begin{table*}
\renewcommand{\arraystretch}{1.15}
\begin{center}
 \begin{tabular}{| c | c | c |c| c |} 
 \hline
 Bin & $\hat{E}$ energy range (TeV) & $\sqrt{TS}$ & Median energy (TeV) & Flux (TeV cm$^{-2}$ s$^{-1}$) \\
 \hline
 c & 1.00-1.78 & 3.06 & 1.71 & (2.21 $\pm$ 0.71) $\times$ 10$^{-12}$ \\
 d & 1.78-3.16 & 7.05 &  2.69 & (4.11 $\pm$ 0.61) $\times$ 10$^{-12}$ \\
 e & 3.16-5.62 & 7.11 & 4.13 & (3.79 $\pm$ 0.56) $\times$ 10$^{-12}$ \\
 f & 5.62-10.0 & 11.5 & 6.45 & (4.50 $\pm$ 0.42) $\times$ 10$^{-12}$ \\
 g & 10.0-17.8 & 16.8 & 10.84 & (4.74 $\pm$ 0.34) $\times$ 10$^{-12}$ \\
 h & 17.8-31.6 & 17.6 & 19.39 & (4.39 $\pm$ 0.34) $\times$ 10$^{-12}$ \\
 i & 31.6-56.2 & 10.6 & 34.59 & (2.29 $\pm$ 0.29) $\times$ 10$^{-12}$ \\
 j & 56.2-100 & 8.24 & 59.16 & (1.77 $\pm$ 0.31) $\times$ 10$^{-12}$ \\
 k & 100-177 & 6.31 & 102.4 & (1.50 $\pm$ 0.35) $\times$ 10$^{-12}$ \\
 l & 177-316 & 0.33 & 131.8 & $<$ 2.74 $\times$ 10$^{-13}$\\

 \hline
\end{tabular}
 \caption{Flux points for eHWC J2019+368. This source is fit to a log parabola. Uncertainties are statistical only. The last bin is not significantly detected so a 95$\%$ CL upper limit is set.}
\end{center}
\end{table*}

\section{Nearest TeVCat sources to each $>$ 100 TeV source}
There are many sources in the TeVCat catalog (tevcat.uchicago.edu) within the region of interest (3$^{\circ}$ radius) for each spectral fit. These sources are listed in Table S8, along with the angular distance between the TeVCat source and the eHWC source. Note that not all TeVCat sources are detected by HAWC. 

\begin{table*}
\renewcommand{\arraystretch}{1.15}
\begin{center}
 \begin{tabular}{| c | c | c | c|} 
 \hline
 Source & TeVCat source & TeVCat source coordinates (RA$^{\circ}$, Dec$^{\circ}$) & Angular distance between \\
 & & & eHWC and TeVCat source ($^{\circ}$)\\
 \hline
 eHWC J1825-134 & 2HWC J1825-134 & (276.46, -13.40) & 0.07 \\
 - & HESS J1826-130 & (276.50, -13.03) & 0.35 \\
 - & HESS J1825-137 & (276.45, -13.78) & 0.41 \\ 
 - & LS 5039 & (276.56, -14.83) & 1.46 \\
 - & 2HWC J1819-150* & (274.83, -15.06) & 2.31 \\
 - & SNR G015.4-00.1 & (274.52, -15.47) & 2.82 \\
 \hline
  eHWC J1907+063 & MGRO J1908+063 & (287.98, 6.27) & 0.08 \\
   - & 2HWC J1908+063 & (287.05, 6.39) & 0.17 \\
  - & SS433 w1 & (287.65, 5.04) & 1.48 \\
   - & 2HWC J1902+048* & (285.51, 4.86) & 2.02 \\
 - & SS433 e1 & (288.40, 4.93) & 2.04 \\
 - & 2HWC J1907+084* & (286.79, 8.50) & 2.18 \\
 - & W49B & (287.78, 9.16) & 2.97 \\
 \hline
  eHWC J2019+368 & 2HWC J2019+367 & (304.94, 36.8) & 0.02 \\
 - & VER J2019+368 & (304.85, 36.80) & 0.10 \\
 - & MGRO J2019+37 & (304.65, 36.83)  & 0.31 \\
 - & VER J2016+371 & (304.01, 37.20) & 1.03 \\
 - & MilagroDiffuse & (305.00, 38.00)  & 1.22 \\
 \hline
 \end{tabular}
 \caption{TeVCat sources within 3$^{\circ}$ of each eHWC that emits above 100 TeV. Note that the Milagro diffuse emission was reported over an extremely large spatial extent; the coordinates listed are the TeVCat source coordinates for this source. }
 \end{center}
 \end{table*}

\clearpage
\bibliography{bib}

\end{document}